\documentstyle[11pt]{article}
\title{\bf Subthreshold photoproduction of charm}
\author{M.A.Braun \thanks{Permanent address: 
Dep. High-Energy physics,
S.Petersburg University, 198504 S.Petersburg, Russia}
\hspace{2 mm} and B.Vlahovic\\
North Carolina Central University, Durham, NC, USA}
\date{}
\def\beq{\begin{equation}}
\def\eeq{\end{equation}}

\begin{document}
\maketitle
\medskip
\vspace{1 cm}

{\bf Abstract}
\vspace{1 cm}

Charm photoproduction rates off nuclei below the nucleon threshold are estimated
using the phenomenologically known structure functions both for $x<1$ and $x>1$.
The rates rapidly fall below the threshold from values  $\sim$10 pb for Pb
close to the threshold (at 7.5 GeV) to $\sim$1pb at 6 GeV.

\section{Introduction}
In view of the envisaged upgrade of the CEBAF facility up to 12 GeV
it becomes important to have relatively secure predictions about the
production rates of charm on nuclear targets below  
the threshold for the nucleon target. This note aims at such
predictions. From the start it has to be recalled that the dynamical
picture of charm production at energies close to the threshold is
much more complicated than at high energies.
On the one hand, the production rates cannot be described
by the standard collinear factorization expression but involve 
gluon distributions both in $x$ and transverse momentum, which are
unknown at small transverse momenta (in the confinement region).
On the other hand, in the immediate vicinity of the threshold
the simple photon-gluon fusion mechanism of charm production becomes
overshadowed by multiple gluon exchanges ~\cite{bro} and formation of colourless
bound states with lower mass, as compared to open charm ~\cite{kop}.
As a result, theoretical studies of the charm
production near the threshold are known to be notoriously difficult.
To avoid all these complications we study charm production 
at not too small distance from the thresholds, where all
effects due to multiple gluon exchanges are hopefully small.
To be more concrete, taking the charmed quark mass $m_c=1.5$ GeV,
we have the threshold of open charm photoproduction on the proton target
at the incident  photon energy $E^{th}_1=7.78$ GeV and on the deuteron
target at 5.39 GeV. 
As we shall argue (see Appendix), our
treatment is hopefully valid at incident photon energies $E$
excluding regions at distances $\sim 0.3$ GeV from these thresholds, say,
$5.7<E<7.5$ GeV.
 Closer to the thresholds 
multigluon exchanges and bound state formation may change our
predictions considerably.

In fact all estimates show that the ratio of the double to single gluon
exchange contributions is determined by the parameter
$\alpha_s(M^2)[\mu/(m_c\Delta x)]^2$, where $\mu$ is the light quark mass, 
$M$ is the mass scale for the gluon coupling to the charmed system and
$\Delta x$ is
the distance from the threshold for
the scaling variable of the produced charmed system (~\cite{bro}, see
also
Appendix). Obviously as $\Delta x\to 0$ multiple gluon exchanges
begin to play the dominant role. They also bind the produced charmed
quarks into colourless bound states, thus changing the threshold value.
However, with $\Delta x$ growing their contribution becomes suppressed.
Taking the effective $\mu\simeq\Lambda_{QCD}=0.3$ GeV, and $M=2m_c$
so that $\alpha_s(M^2)\sim 0.3$ we
find the suppression factor of the order $75(\Delta x)^2$. This factor
determines the region where one can neglect all multiple gluon
exchanges and bound state formation. In a more detail this 
suppression factor will be discussed in Appendix.

Neglecting multiple gluon exchanges, subthreshold production 
will be determined by the gluon distribution in the "cumulative"
region ($x>1$ at large energies or massless target). As mentioned,
in the low energy region the charm production rate involves 
the gluon distribution not only in $x$ (collinear factorization)
but also in both $x$ and $k_{\perp}^2$
($k_{\perp}$ factorization). We shall use the simplest approximation 
about the structure of this combined distribution, assuming 
the dependence on $x$ and $k_{\perp}$ factorized.
 For the latter dependence we
shall use a monopole form, prompted by the perturbation theory.
As for the former one, we shall exploit the existing (scarce) data
on the nuclear structure functions at $x>1$, from which we shall
extract the relevant gluon distribution, using the DGLAP evolution
equation.
 With all these simplifications, we
hope to be able to predict the rates up to factor 2$\div$3 \footnote
{Some crude estimates were earlier reported in ~\cite{BV}. Their order
agrees with the present calculations at energies not too close to the
threshold.}.

\section{Kinematics and cross-sections}
\subsection{The charm production cross-section}
Consider the exclusive process 
\beq
\gamma+A\to C{\bar C}+A^*,
\eeq
where $A$ is the target nucleus of mass $m_A$ and
$A^*$ is the recoil nuclear system of mass $m_A^*$. 
We denote the total mass of the
$C\bar C$ system as $M$.
Obviously $M\geq 2m_c$ where $m_c$ is the mass of the C-quark, which 
we take as 1.5 GeV. The inclusive cross-section
for charm photoproduction is obtained after summing over all states of the
recoil nuclear system.

We choose a reference system in which the target nucleus 
with momentum $Ap$ is at rest and the incoming photon with momentum $q$ is
moving along the $z$-axis in the opposite direction, so that $q_+=q_{\perp}=0$. 
The  photoproduction cross-section corresponding to (1) is then obtained via the
imaginary part of the diagram in Fig. 1 as
\beq
\sigma_{A\to A^*}=A\int\frac{d^4k}{(2\pi)^3}\delta((Ap-k)^2-{m_A^*}^2)x
\left(\frac{\Gamma_{AA^*}(k^2)}{k^2}\right)^2\sigma_g(M^2,k^2).
\eeq
Here $\Gamma_{AA^*}(k^2)$ is the vertex for gluon emission from the
target; 
$\sigma_g(M^2,k^2)$
is the photoproduction cross-section off the virtual gluon of momentum $k$.
We have also introduced the scaling variable for the gluon as $x=k_+/p_+$. 
Due to
$q_+=0$ this is also the scaling variable for the observed charm. Note that 
this
definition, which is standard at large energies and produced masses, is
not 
at all standard
at moderate scales. In particular this $x$ does not go to unity at the 
threshold
for the nucleon target. Rather the limits for its variation converge 
to a common
value 0.76. For the nuclear targets with $A>>1$ its minimal value at the 
nucleon threshold is 
well below unity ($\sim$ 0.64). One should have this in mind when
associating this $x$
with the gluon distribution: it follows that for a nuclear target, for 
energies going 
noticeably below the nucleon threshold, the 
cumulative region (prohibited for the free nucleons
kinematics) includes not only values of $x$ above unity but
also a part of the region $x<1$.

We use the $\delta$-function to integrate over $k_-$ to obtain
the cross-section (2) as
\beq
\sigma_{A\to A^*}=A\int\frac{dxd^2k_{\perp}}{2(A-x)(2\pi)^3}
\left(\frac{\Gamma_{AA^*}(k^2)}{k^2}\right)^2\sigma_g(M^2,k^2).
\eeq
In these variables we find
\beq
M^2=xs_1+k^2,\ \ s_1=2pq
\eeq
and
\beq
k^2=xAm^2-\frac{x}{A-x}{m_A^*}^2-\frac{A}{A-x}k^2_{\perp},
\eeq
where we have put $p^2=m^2$, the nucleon mass squared, neglecting the
binding.

The limits of integration in (3) are determined by the condition $M^2\geq
4m_c^2$,
 which leads to
\beq
x(s_1+Am^2)-\frac{x}{A-x}{m_A^*}^2-\frac{A}{A-x}k^2_{\perp}-4m_c^2\geq 0.
\eeq
Since $k_{\perp}^2\geq 0$, one gets
\beq
x(s_1+Am^2)-\frac{x}{A-x}{m_A^*}^2-4m_c^2\geq 0,
\eeq
from which one finds the limits of integration in $x$ for the transition
$A\to A^*$:
\beq
x_1^{A\to A^*}\leq x\leq x_2^{A\to A^*}
\eeq
where
\beq
x_{1,2}^{A\to A^*}=\frac{1}{2s}\Big(As-{m_A^*}^2+
4m_c^2\pm\sqrt{[As-(m^*_A+2m_c)^2][As-(m^*_A-2m_c)^2]}\Big)
\eeq
and $s=s_1+Am^2$.
The limits of integration in $k_{\perp}$ at a given $x$ are determined by
(6).

Using (5) we may pass from the integration variable $k_{\perp}^2$ to
$|k^2|$. Summing over all states of the recoiling nucleus $A^*$ we get
\beq
\sigma_A=\int_{x_1^{(A)}}^{x_2^{(A)}}
xdx\int_{|k^2|_{min}}^{xs_1-4m_c^2}
d|k^2|
\sigma_g(xs_1-|k^2|,k^2)\rho(x,|k^2|),
\eeq
where
\beq
\rho(x,|k^2|)=\frac{\pi}{2(2\pi)^3}
\sum_{A^*}\left(\frac{\Gamma_{AA^*}(k^2)}{k^2}\right)^2,
\eeq
\beq
|k^2|_{min}=\frac{A}{A-x}x^2m^2
\eeq
and $x_{1,2}^{(A)}$ is determined by (9) with $m_A^*$ put to its minimal
value
$m_A^*=m_A=Am$. 

The threshold energy corresponds to $x_1^{(A)}=x_2^{(A)}$ or
$As=(Am+2m_c)^2$. In 
terms of the photon energy $E$ we have $s_1=2mE$ and the threshold energy
is found to
be
\beq
E^{th}_A=2m_c\left(1+\frac{1}{A}\frac{m_c}{m}\right).
\eeq
It steadily falls with $A$ from the nucleon target threshold. With
$m_c=1.5$  GeV we find (in GeV):
\beq
E^{th}_1=7.79,\ \ E^{th}_2=5.39,\ \ E^{th}_3=4.60,\ \ E^{th}_{12}=3.40,\ \ 
E^{th}_{207}=3.02
\eeq
down to the value $2m_c=3$ GeV for infinitely heavy nucleus.

\subsection{High-energy limit}
To interprete $\rho$ in Eq. (10) it is instructive to study
its high-energy limit, which corresponds to taking  $s_1 >>m_c^2$ and both 
quantities much greater than
the nucleon mass. Assuming that the effective values of the gluon
virtuality are limited (and small) one then gets for the nucleon target ($A=1$)
\beq
\sigma_1= 
\int_{4m_c^2/s_1}^{1}xdx\sigma_g(xs_1)
\int_{0}^{xs_1}d|k^2|\rho(x,|k^2|).
\eeq
Here we also neglect the off-mass-shellness of the cross-section off the 
gluon, considering
$|k^2|<<4m_c^2$.
The obtained formula is precisely the standard collinear factorization formula
with the identification
\beq
xg(x,M^2)=\int_{0}^{M^2}d|k^2|x\rho(x,|k^2|).
\eeq
Thus the quantity $\rho(x,|k^2|)$ obviously has a meaning of the double
distribution of gluons in $x$ and $|k^2|$.

\section{The gluon distribution $\rho(x,|k^2|)$}
To find the double distribution of gluons in $x$ and $|k^2|$ one may be tempted
to use (16) and simply differentiate $xg(x,M^2)$ in $M^2=|k^2|$. However
(16) is only true for $x<<1$. At finite $x$ the derivative $dg(x,m^2)/dM^2$
is not positive and cannot be interpreted as the double gluonic distribution.

To avoid this problem, we choose a different, somewhat simplified approach. 
We assume a simple factorizable form for the double 
density
$\rho(x,|k^2|)$ and choose the $|k^2|$ dependence in accordance with the 
perturbation theory,
with an infrared cutoff in the infrared region:
\beq
\rho(x,|k^2|)=\frac{a(x)}{|k^2|+\Lambda^2}.
\eeq
Function $a(x)$ can be obtained matching (17) with the observed
$xg(x,M^2)$ at a particular point
$M_0^2$. Since we are interested in the threshold region, we take 
$M_0=2m_c$ to finally obtain
\beq
\rho(x,|k^2|)=\frac{g(x,4m_c^2)}{\ln(4m_c^2/\Lambda^2+1)}
\ \frac{1}{|k^2|+\Lambda^2}.
\eeq
The recipe (18) amounts to taking in (16) $\rho$ dependent also on $M^2$,
with the latter dependence factorized.
Our calculations show that the results are rather weakly  dependent of the
infrared cutoff chosen in the interval 0.4$\div$0.7 GeV.

For the proton at $x<1$ the gluon distribution $g(x,4m_c^2)$ can be taken from 
numerous existing fits to the experimental data. In our calculations we
have used GRV95 LO ~\cite{GRV}. For the nucleus in the non-cumulative region
\beq
x_1^{(1)}<x<x_2^{(1)}
\eeq 
we use the simplest assumption $g_A(x,Q^2)=Ag_1(x,Q^2)$ 
neglecting the EMC effect in the first
approximation. Obviously this not a very satisfactory  approximation at
$x$ quite close to the threshold for the proton target. However our
estimates are in any case not justified in this region, since, as
mentioned in the introduction, multiple gluon exchanges and bound states 
formation begin to play a leading role in the immediate vicinity of the
threshold. 

For the nuclei in the cumulative region (outside region (19)) the gluon
distribution
may be estimated using, first, the existing data for the 
nuclear structure functions in this region
and, second, the popular hypothesis that at sufficiently low $Q^2=Q_0^2$
the sea and gluon  distributions vanish and hadrons become 
constructed exclusively of valence quarks.
Then one can find the gluon distribution at a given $Q^2$
from the standard DGLAP evolution equation, with the quark distributions
determined from the experimental data on the structure functions at $x>1$
and evolved back to $Q^2=Q_0^2$.  In practice we took the initial
valence distributions  in carbon at $Q^2=Q_0^2$ in the form
\beq
u(x,Q_0^2)=d(x,Q_0^2)=a\,e^{-bx}
\eeq
and the rest of the distributions equal to zero. 
Then we calculated the carbon structure function  at $x>1$ 
and $Q^2$ in correspondence with the data of ~\cite{exp} and chose the
parameters $a$ and $b$ to fit the data. With thus chosen $a$ and $b$
we finally calculated the gluon distribution in carbon at the scale
$4m_c^2$. Our obtained gluon distributions in carbon for $Q_0=0.4$ and 0.7
GeV/c are shown in Fig. 2 for $1<x<2$. As one observes, the
dependence on the choice of $Q_0$ is very weak in this interval. 
The slopes result equal to 11.4 ($Q_0=0.4$ GeV/c) and 11.2 ($Q_0=0.7$).

The distribution for other nuclei was taken from the
A-dependence, chosen in accordance with the experimental data for
hadron production at $x>1$ as $\propto A^{1+0.3x}$ ~\cite{cum}.

\section{Numerical results}
The cross-section (10) involves the photon-gluon fusion cross-section
$\sigma_g$ off mass shell.  The integration over the gluon virtuality starts
from $|k^2|\sim m^2$. If one assumes $m/M\to 0$ then the bulk
of the contribution
will come from the region of small $|k^2|$ (with a logarithmic precision).
In reality $m/M$ is not so small. However, to simplify our calculations,
as a first approximation, we have taken the photon-gluon cross-section on 
the mass shell,
where it is known to be [5]
\beq
\sigma_g(M^2)=\pi\alpha_{em}\alpha_s e_c^2\frac{1}{M^4}
\int_{t_1}^{t_2} dt\Big[\frac{t}{u}+\frac{u}{t}+\frac{4m_c^2M^2}{tu}
\left(1-\frac{m_c^2M^2}{tu}\right)\Big]
\eeq
Here $e_c$ is the quark charge in units $e$, $u=-M^2-t$ and the limits
$t_{1,2}$ are given by
\beq
t_{1,2}=-\frac{M}{2}[M\pm\sqrt{M^2-4m_c^2}]
\eeq
We take the strong coupling constant $\alpha_s=0.3$.

Our gluon distribution depends on  two parameters: the  infrared cutoff
$\Lambda$ in (19) and the value of $Q_0$, at which the sea and gluon
distributions die out. The order of both is well determined, but still one
can vary them to some degree. In our calculations we took both $\Lambda$
and $Q_0$ equal to 0.4 or 0.7 GeV/c. 
 
With these values for the parameters we obtain the cross-sections for charm 
photoproduction on Pb shown in Fig. 3. As one observes, the
dependence on both  $\Lambda$ and $Q_0$ is relatively weak:
 in the whole range of their variation the cross-sections change by less 
than 30\%.
Fig. 4 illustrates the A-dependence of the cross-sections (with
$\Lambda=Q_0=0.4$ GeV/c). 
To have the idea of the number of nucleons which have to 
interact
together to produce charm at fixed energy below threshold we show the limits of
intergation $x_1$ and $x_2$ in Fig. 5.

As expected, the cross-sections rapidly fall for energies below threshold.
Their energy dependence cannot be fit with a simple exponential (in fact
they fall faster than the exponential). As to the aboslute values, for Pb
the cross-section fall from $\sim$ 10 pb immediately below the 
threshold down to $\sim$ 1 pb at $E=6$ GeV.
The $A$ dependence is close to linear.

\section{Discussion}
We have  estimated charm photoproduction rates for nuclear
targets below  the nucleon target threshold.
The estimates require knowledge of the gluon distribution in both $x$ and 
$k_{\perp}^2$ in a wide region of the momenta including the confinement region.

Our estimates were based on a simple factorization assumption and introduction
of an infrared cutoff. Another approximation  has been to take the 
photon-gluon fusion cross-section on its mass-shell. 
We are of the opinion that this
second approximation is not very serious, as compared to the first one.
In any case it can easily be dropped for the price of considerable
complication of the calculation.

Our predictions are infared cutoff dependent. However the cutoff
dependence results weak for variations of the cutoffs in a reasonable
interval.

In our study we assumed the standard mechanism of charm production via 
gluon-photon fusion (a single gluon exchange between light and heavy quarks).
It can be shown that this mechanism dominates, provided one is not too
close to the threshold (see Appendix).
  
\section{Acknowledgements}

M.A.B. is thankful to the Faculty of Science of the NCCU
for hospitality. 

\section{Appendix. Multiple gluon exchange}
\subsection{Kinematics and phase volume}
To study the relative weight of multiple gluon exchange  
 we consider a simplified picture with
a mesonic target composed of a light quark and antiquark of mass $\mu$.
We neglect the binding, so that the meson mass is just $2\mu$. We shall 
compare  contributions to heavy flavour production of the three
amplitudes Fig. 6$a-c$. Amplitude $a$ corresponds to a single
gluon exchange between light and heavy quarks, amplitudes $b$ and $c$
to double gluon exchange. 
We use the light-cone variables and denote 
$
k_{i+}=z_ip_+, \ \ p_{i+}=x_ip_+,\ \ i=1,2.
$

The phase volume for the reaction is given by
\beq
dV=\frac{1}{16(2\pi)^8}\frac{d^3k_1d^3p_1d^3p_2}{z_1z_2x_1x_2}\delta(R_e-R),
\eeq
where $d^3k_1=dz_1d^2k_{1\perp}$ etc and the $\delta$-function
refers to conservation of the light-cone energy (the "-" component
of the momentum). Its argument contains the external energy
$
R_e=2pq+4\mu^2=2\mu E+4\mu^2,
$
and the energy of the produced particles
$
R=\sum_{i=1}^2(m_{c,i\perp}^2/z_i+
\mu_{i\perp}^2/x_i)
$
The minimal value of $R$ determining the production threshold occurs at 
\beq
z_i=z_0=\frac{m_c}{m_c+\mu},\ \ x_i=x_0=\frac{\mu}{m_c+\mu},\ \
k_{i\perp}=p_{i\perp}=0,
\ \ i=1,2,
\eeq
and is equal to
$
\min\, R=R_0=2(m_c+\mu)^2.
$

We shall study our amplitudes near the threshold, so that $x_i$ will be
small and $z_i$ will be close to unity. We put
$
z_i=z_0+\zeta_i,\ \ x_i=x_0+\xi_i,\ \ i=1,2 
$
and develop $R$ near the threshold keeping terms of the second order
in $\zeta$'s,  $\xi$'s and transverse momenta. 
We present the difference $R_e-R$ in the form
$
R_e-R_0=\Delta R_0,
$
where dimensionless $\Delta$ measures the distance from the threshold
and is supposed to be small.
Finally we rescale our variables as follows
\beq
\zeta_i=\tilde{\zeta}_i\sqrt{\Delta\frac{z_0^3R_0}{M^2}},\ \
\xi_i=\tilde{\xi}_i\sqrt{\Delta\frac{x_0^3R_0}{m^2}},\ \
k_{i\perp}=\tilde{k}_{i\perp}\sqrt{\Delta z_0R_0},\ \
p_{i\perp}=\tilde{p}_{i\perp}\sqrt{\Delta x_0R_0}.
\eeq
Obviously new variables with tildes are dimensionless and of the
order unity. Using this,  one finds that approximately
$\tilde{\zeta}_1=-\tilde{\zeta}_2$ and $\tilde{k}_2=-\tilde{k}_1$. 
So
the phase volume acquires the form 
\beq
dV=V_0\Delta^{7/2}\frac{d\tilde{z}_1d\tilde{\xi}_1d\tilde{\xi}_2
d^2\tilde{k}_{1\perp}
d\tilde{p}_{1\perp}d\tilde{p}_{2\perp}}
{(x_0+\tilde{\xi}_1\sqrt{2\Delta x_0})(x_0+\tilde{\xi}_2\sqrt{2\Delta x_0})}
\delta\Big(1-2\tilde{\zeta}_1^2
-\tilde{\xi}_1^2-\tilde{\xi}_2^2
-2\tilde{k}_{1\perp}^2
-\tilde{p}_{1\perp}^2-\tilde{p}_{2\perp}^2\Big),
\eeq
where
$
V_0=m^3M\sqrt{z_0/2}/(2\pi)^8.
$

\subsection{Amplitudes}

We  use the Coulomb gauge for the interaction between quarks
and neglect  the contribution from the transverse momenta in it. Then the
interaction depends only on the scaling variables, and for the transition
between, say, light quarks $p_1+p_2\to p'_1+p'_2$ is given by
\beq
V(p_1,p_2|p'_1,p'_2)=4\pi\alpha_s\frac{(x_1+x'_1)(x_2+x'_2)}{(x_1-x'_1)^2}.
\eeq
We shall assume that the initial light quarks have their momenta equal to $p$,
so that their scaling variable is equal to unity. We omit the the common factor
due to their binding into the initial target meson.
Finally we consider photoproduction, so that $q^2=0$ and choose a system
in which $q_+=q_{\perp}={\bf p}=0$.

The amplitude corresponding to Fig. 6$a$ is given by
\beq
{\cal A}^{(a)}=\frac{V((p,p|2p-p_2,p_2)V(q-k_1,2p-p_2|k_2,p_1)}
{(\mu^2-(2p-p_2)^2)(m_c^2-(q-k_1)^2)}.
\eeq
The two interactions near the threshold turn into
$ 12\pi\alpha_s$ and $2\pi\alpha_s(z_2-z_1)$
where we used the fact that $x_1,x_2<<1$ and $z_1,z_2\simeq 1$.
For the same reason we find the two denominators as
$
\mu^2-(2p-p_2)^2\simeq
2\mu_{2\perp}^2/x_2
$
and
$
m_c^2-(q-k_1)^2\simeq 2pq
$.
In our dimensionless variables we obtain
 \beq
{\cal A}^{(a)}=-2c_1\sqrt{2z_0\Delta}\frac{\tilde{\zeta}_1
(x_0+\tilde{\xi}_1\sqrt{2x_0\Delta})}
{\mu^2+2\mu(m_c+\mu)\Delta\tilde{p}_{2\perp}^2},\ \
c_1=\frac{6\pi^2\alpha_s^2}{pq}. 
\eeq

The  amplitude corresponding to Fig. 6$b$ is
\beq
{\cal A}^{(b)}=\frac{V((q-k_1,p|q-k_1+p-p_1,p_1)V(q-k_1+p-p_1,p|k_2,p_2)}
{(m_c^2-(q-k_1)^2)(m_c^2-(q-k_1+p-p_1)^2)}.
\eeq
Near the threshold the interactions become $-4\pi\alpha_s$ and
$4\pi\alpha_s$
The new denominator is
\beq
m_c^2-(q-k_1+p-p_1)^2\simeq
m_c^2+2(m_c+\mu)\Delta(\sqrt{m_c}\tilde{k}_1+\sqrt{\mu}\tilde{p}_1)_{\perp}^2.
\eeq
At $\Delta<<1$ we can drop the second term. So we get
\beq
{\cal A}^{(b)}=-c_2\frac{1}{m_c^2},\ \ c_2=\frac{8\pi^2\alpha_s^2}{pq}.
\eeq

Finally we consider the amplitude  of Fig. 6$c$:
\beq
{\cal
A}^{(c)}=\frac{V((k_1-p+p_1,p|k_1,p_1)V(k_2-p+p_2,p|k_2,p_2)}
{(m_c^2-(k_1-p+p_1)^2)(m_c^2-(k_2-p+p_2)^2)}.
\eeq
Near the threshold both interactions become approximately 
equal to $4\pi\alpha_s$ and
both denominators to $m_c^2$. So the amplitude becomes
\beq
{\cal A}^{(c)}=2c_2\frac{pq}{m_c^4}\simeq 2c_2\frac{1}{m_c^2},
\eeq
where we have used that near the threshold $pq\simeq m_c^2$.
So the amplitudes $b$ and $c$ have the same order of magnitude.

\subsection{Cross-sections}
Now we can pass to our main goal: comparison of contributions of 
the three amplitudes to the
total cross-section for heavy flavour production.  The first thing to note is 
that near the threshold amplitude $a$ does not interfere with $b$ and $c$, since
since ${\cal A}^{(a)}$ is odd in $\zeta_1$ and 
${\cal A}^{(b,c)}$ do not depend on $\zeta_1$ at all. Second, since 
${\cal A}^{(b)}$
and ${\cal A}^{(c)}$ are of the same order and structure it is sufficient
to compare
the contributions of ${\cal A}^{(a)}$ and ${\cal A}^{(b)}$. Finally, due to
the fact
that $x_0$ is small, the magnitude of contributions depends on the 
relation between $\Delta$ and $x_0$. We shall study two limiting cases:
$\Delta<<x_0$ (region A) and $\Delta >>x_0$ (region B).

Region A refers to the production immediately above the threshold. 
In
this case
we can take $x_{1,2}\simeq x_0$. Then we find the contribution of
amplitude $a$ to the cross-section as
\beq
\sigma^{(a)}=V_0\Delta^{7/2}\frac{1}{x_0^2}
[2c_1\sqrt{2\Delta}\frac{x_0}{\mu^2}^2]\, I^{(a)},
\eeq
where $I^{(a)}$ is a certain integral of the order unity.
The contribution of the amplitude $b$ to the cross-section will be
\beq
\sigma^{(b)}=V_0\Delta^{7/2}\frac{1}{x_0^2}[c_2\frac{1}{m_c^2}]^2\,
I^{(b)},
\eeq
 where $I^{(b)}$ is another integral of the order unity.
The ratio of these two cross-sections will have the order
\beq
\frac{\sigma^{(b)}}{\sigma^{(a)}}
\sim\frac{\mu^2(m_c+\mu)^2}{m_c^4\Delta}\sim\frac{\mu^2}{m_c^2\Delta}
\sim\frac{x_0^2}{\Delta}.
\eeq
Thus immediately above the threshold the contribution from  
amplitudes $b$ and $c$ dominate.
However with the growth of $\Delta$, in the region $x_0^2<<\Delta$ 
(and $<<x_0$ to still remain
in  region A) the contribution of  amplitudes of $b$ and $c$ become
suppressed by factor $\mu/m_c$.

In region B we can approximate
$x_{1,2}\simeq\tilde{\xi}_{1,2}\sqrt{2x_0\Delta}$.
To avoid logarithmic divergence in $\tilde{\xi}_{1,2}$ 
we cutoff the integration region from below at values of the order 
$\sqrt{x_0/\Delta}$. We also note that
the integral over $\tilde{p}_{1\perp}$ appearing in the contribution of
amplitude $a$
is well convergent at values of $\tilde{p}_{1\perp}^2\sim x_0/\Delta$ 
so that we may neglect $\tilde{p}_{1\perp}^2$ in the argument of the
$\delta$-
function  and separate the integration over $\tilde{p}_{1\perp}$ as a factor
\beq
\int\frac{d^2\tilde{p}_{1\perp}}
{[\mu^2+2\mu(m_c+\mu)\Delta\tilde{p}_{2\perp}^2]^2}=
\frac{\pi}{2\mu^3(m_c+\mu)\Delta}.
\eeq
We find the cross-section from ${\cal A}^{(a)}$ as
\beq  
\sigma^{(a)}=V_0\Delta^{5/2}\frac{1}{2x_0}[2c_12\Delta
\sqrt{x_0}]^2\frac{\pi}{2\mu^3(m_c+\mu)\Delta}\,J^{(a)},
\eeq
where $J^{(a)}$ is an integral of the order $\ln (x_0/\Delta)$. 
The cross-section from ${\cal A}^{(b)}$ is found to be
\beq  
\sigma^{(b)}=V_0\Delta^{5/2}\frac{1}{2x_0}[c_2\frac{1}{m_c^2}]^2\,J^{(b)},
\eeq
with $J^{(b)}$ an integral of the order $\ln^2 (x_0/\Delta)$.

The ratio of the two cross-section turns out to be of the same order
up to a logarithmic factor
\beq
\frac{\sigma^{(b)}}{\sigma^{(a)}}\sim\frac{\mu^2(m_c+\mu)^2}{m_c^4\Delta}
\ln\frac{\Delta}{x_0}
\sim\frac{\mu^2}{m_c^2\Delta}\ln\frac{m_c\Delta}{\mu}
\sim\frac{x_0^2}{\Delta}\ln\frac{\Delta}{x_0}
\eeq
and so the contribution of amplitude $a$ clearly dominates in region B,
where $\Delta>>x_0$. The suppression factor for the contribution of the
amplitudes $b$ and $c$ with double gluon exchange is found to be
$m_c^2/(\mu^2\Delta)$. With $m_c/\mu\sim 5$ it is of the order $25\Delta$.
Taking into account that double gluon exchange involves a coupling constant
at the heavy flavour mass scale will add a factor $\sim$3 more. So in the
end we find a suppression factor of the order $75\Delta$, which implies
that at a distance of 0.3 GeV from the threshold the contribution of the
double gluon exchange drops by a factor $\sim$3.

\subsection{Hadrons with more quarks}
Generalization to hadrons with more quarks is straightforward,
although quite cumbersome due to rapid proliferation of diagrams.
However the basic findings remain unchanged. Indeed all the difference
between amplitides with a single and double gluon exchange between
light and heavy quarks comes from the fact that a soft propagator
of the order $x/\mu_{\perp}^2$ in the amplitude with a single gluon
exchange is substituted by a hard propagator of the order $1/(pq)$ in the
amplitude with two gluon exchanges. This difference persists for any
number of quarks in the hadron, although the total number of soft
propagator grows. So although the overall dependence on $\Delta$
will change (in accordance with the quark counting rules), the relation
between cross-sections with a single and double gluon exchange will
remain the same.

\subsection{Bound states}
One may wonder if the production cross-section is dominated by
the formation of final bound states, via diagrams as Fig. 7$a$,
which looks as quark rearrangement without any gluon exchange ~\cite{kop}.
However one has to recall that in the bound state of a light and a heavy
quark (D-meson) the typical configuration requires $p_{i+}/l_{i+}=\mu/m_D$
and $k_{i+}/l_{i+}=m_c/m_D$, where we neglect the binding taking
$m_D=m_c+\mu$. The initial light quarks have however $p_{i+}=p_+$.
So for their binding into D-mesons, they have to diminish their
longitudinal momenta by at least two hard gluon exchanges, as shown in
Fig. 7$b$. But the process in Fig.7$b$ contributes actually a
part of the cross-section generated by the amplitude $A^{(c)}$
studied in the preceding subsection, which corresponds to the 
immediate binding of the open charm into D-mesons. Above the threshold
of the open charm production its contribution can only be
smaller than the
total rate of open charm production. True, immediately below this
threshold, at distances of the order of the binding energy, this mechanism 
is obviously the only one that contributes, in agreement with the
estimates above for very small $\Delta$'s. However, as we have seen, with
the growth of $\Delta$ the strength of multiple  interactions between 
light and heavy quarks necessary to produce them in a state 
appropriate for their binding rapidly goes down. With them goes down also
the correspondiing part of the cross-section due to immediate binding.

%\section{References}

\section{Figure captions}

Fig. 1. The forward scattering amplitude corresponding to reaction 
(1). Heavy quarks are shown by double lines.

Fig. 2. The cumulative ($x>1$) gluon distributions in carbon at 
$Q^2=4m_c^2$. The upper (lower) curve corresponds to $Q_0=0.4(0.7)$ GeV/c.

Fig. 3. The charm photoproduction subthreshold cross-sections off Pb 
for different choice of parameters $\Lambda$ and $Q_0$. Curves from top to
bottom correspond to ($\Lambda,Q_0$)=(0.7,0.4),(0.4,0.4),\\(0.7,0.7) and
(0.4,0.7) GeV/c.

Fig. 4. The charm photoproduction subthreshold cross-sections for
different  targets. Curves from bottom to top correspond to 
$A$ = 12,,64 and 207.

Fig. 5. The limits of $x$-integration for different photon energies and
nuclear targets. Curves from bottom to top correspond to 
$A$ =12, 64 and 207.

Fig. 6. Amplitudes for charm photoproduction off a meson with
a single ($a$) and double ($b,c$) gluon exchanges between light and heavy
quarks. The latter are shown with double lines. Vertical lines correspond
to gluonic exchanges.

Fig. 7. Amplitudes for the $D\bar{D}$ photoproduction off a meson.
Diagram $a$ is equivalent to diagram $b$, which shows how the
produced quarks aquire their momenta appropriate for the binding. 
Notations are as in Fig. 6.
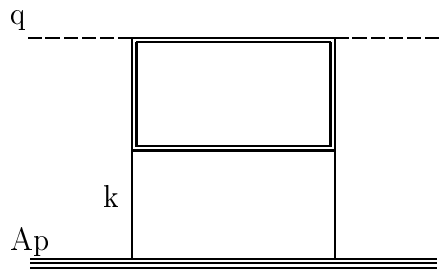
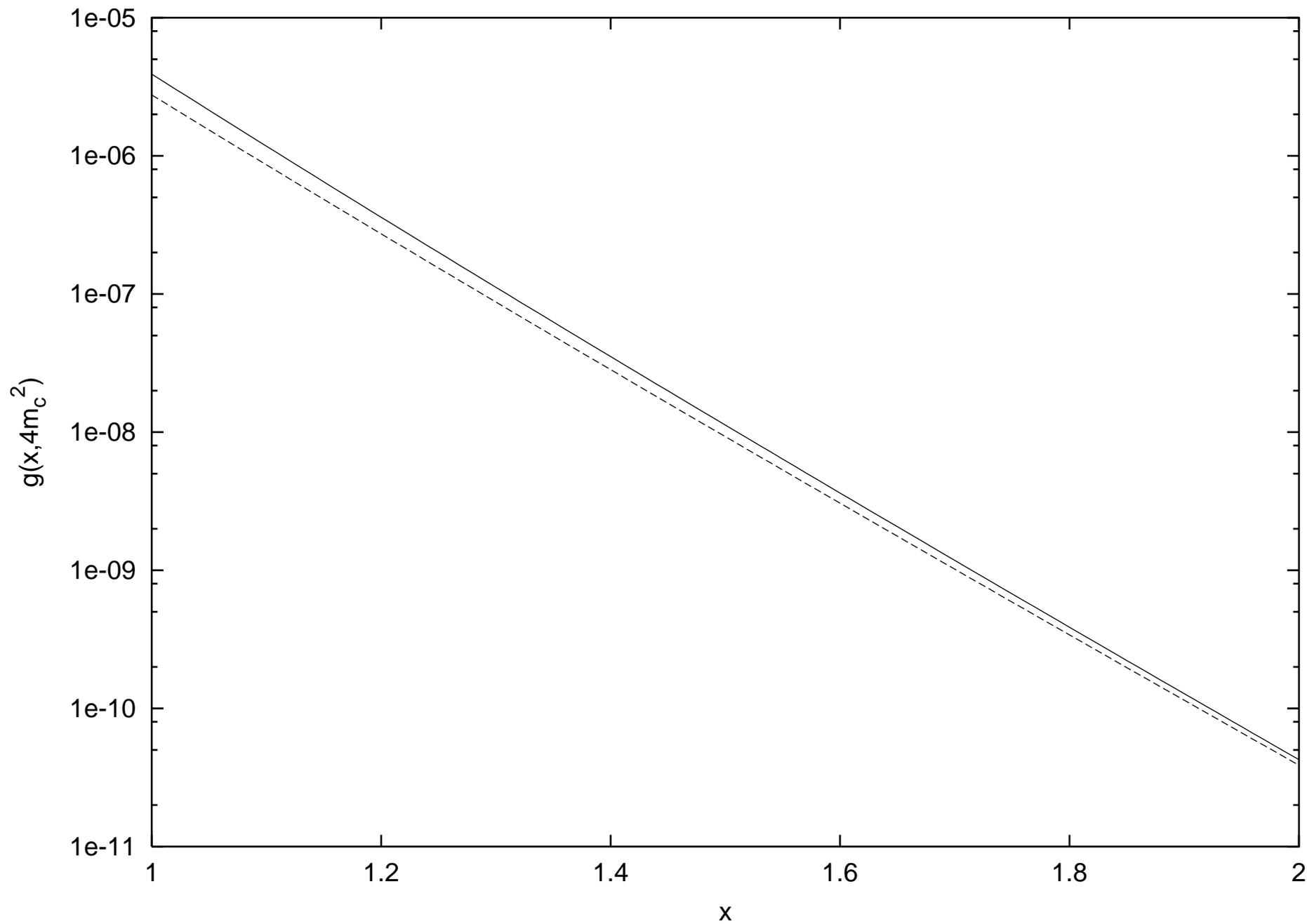
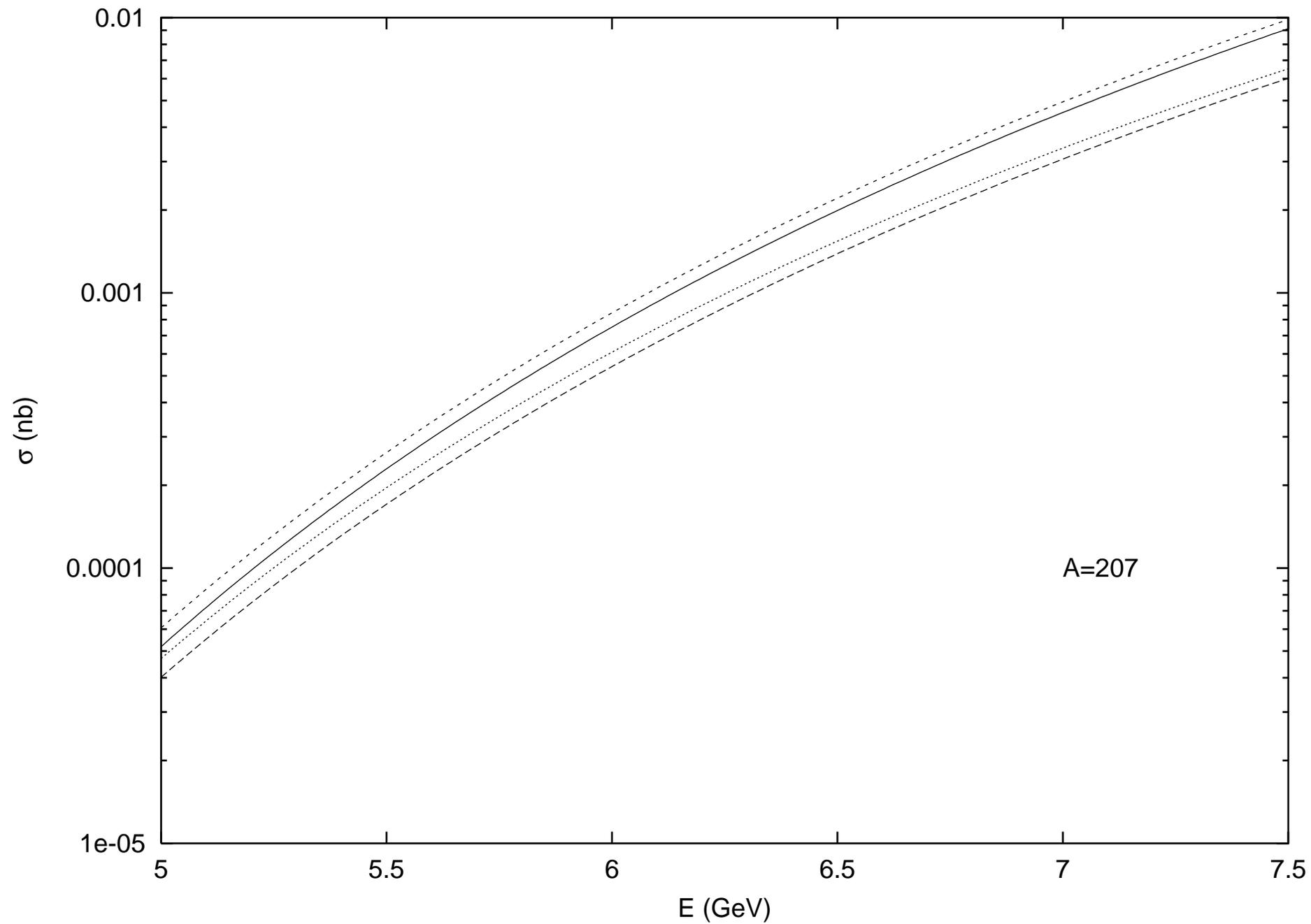
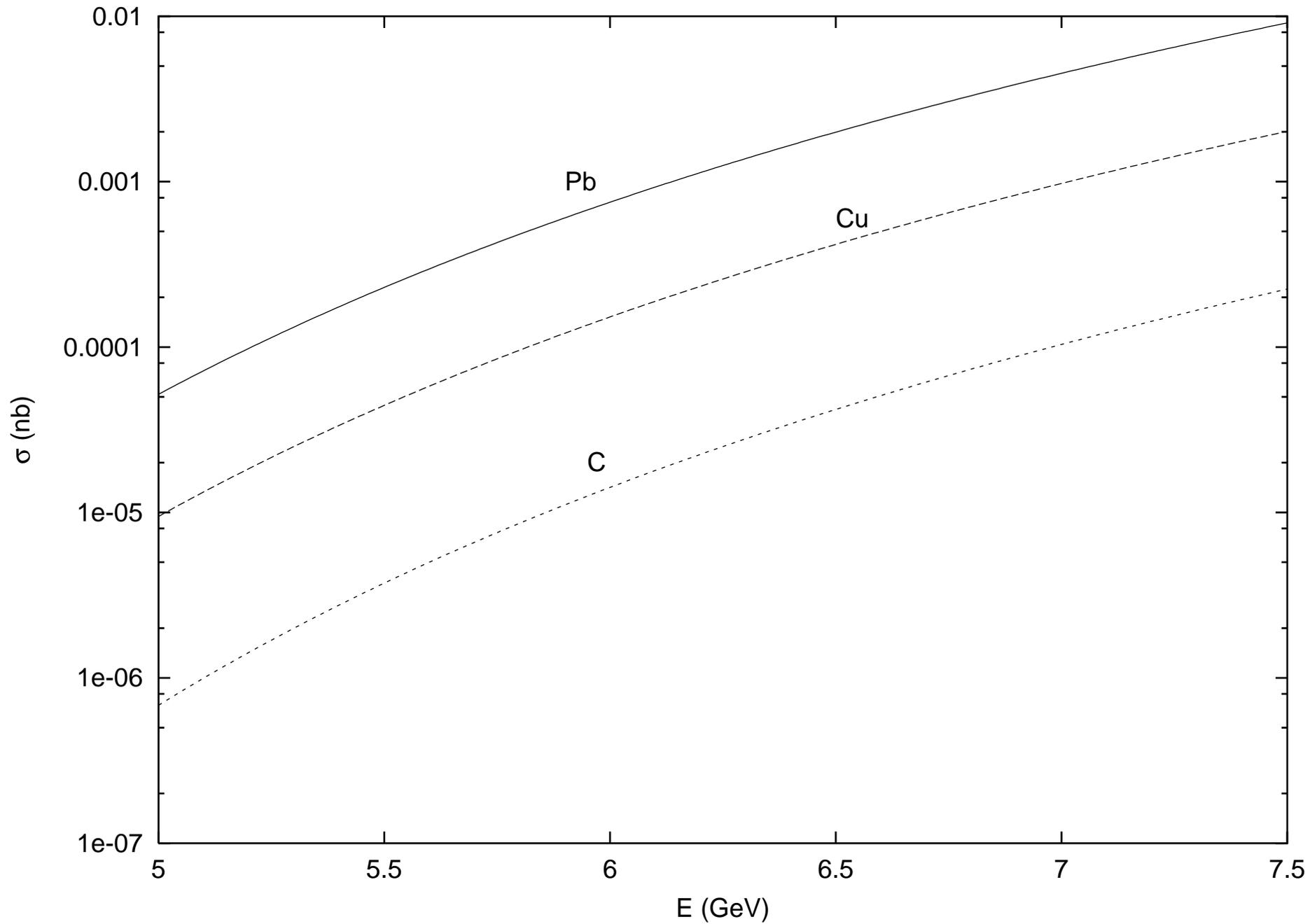
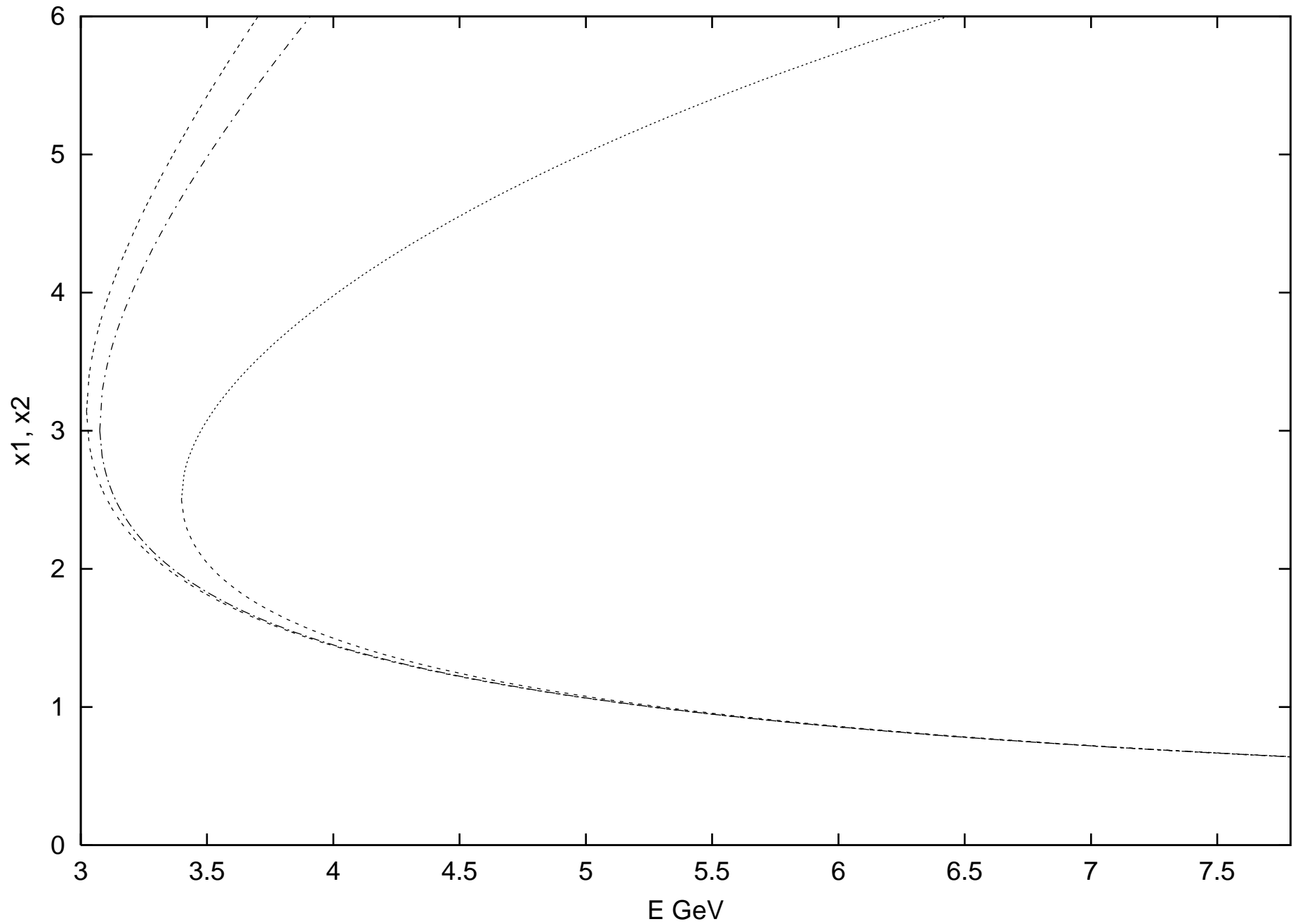
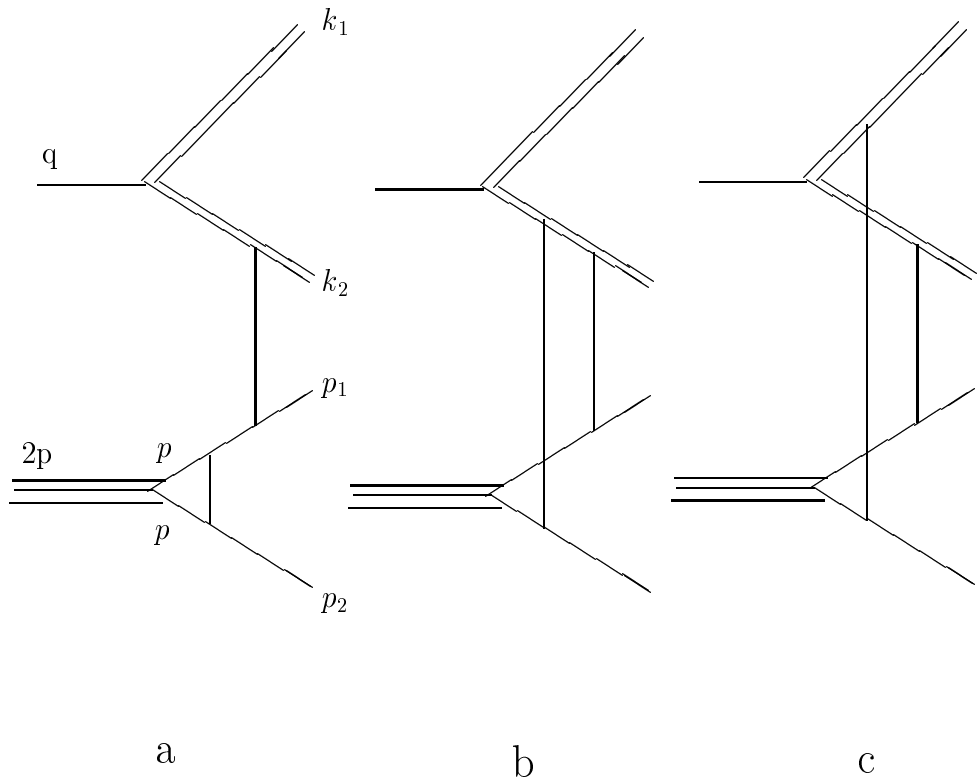
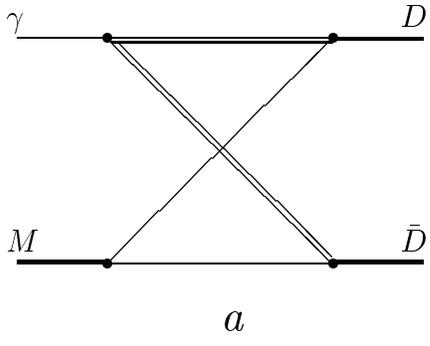
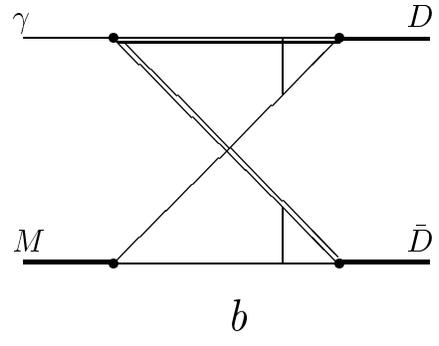
\end{document}